\documentstyle{article}

\def\1ad{\mbox{\normalsize $^1$}}
\def\2ad{\mbox{\normalsize $^2$}}
\def\3ad{\mbox{\normalsize $^3$}}
\def\4ad{\mbox{\normalsize $^4$}}
\def\5ad{\mbox{\normalsize $^5$}}
\def\6ad{\mbox{\normalsize $^6$}}
\def\7ad{\mbox{\normalsize $^7$}}
\def\8ad{\mbox{\normalsize $^8$}}
\def\makefront{\vspace*{1cm}\begin{center}
\def\newtitleline{\\ \vskip 5pt}
{\Large\bf\titleline}\\
\vskip 1truecm
{\large\bf\authors}\\
\vskip 5truemm
\addresses
\end{center}
\vskip 1truecm
{\bf Abstract:}
\abstracttext
\vskip 1truecm}
\def\be{\begin{equation}}       \def\eq{\begin{equation}}
\def\ee{\end{equation}}         \def\eqe{\end{equation}}

\def\bea{\begin{eqnarray}}      \def\eqa{\begin{eqnarray}}
\def\ena{\end{eqnarray}}        \def\eea{\end{eqnarray}}
                                \def\eqae{\end{eqnarray}}

\def\ba{\begin{array}}
\def\ea{\end{array}}
\def\unit{1 \hskip-.3em \raise2pt\hbox{$ \scriptstyle |$ } }

\def\a{\alpha}
\def\b{\beta}
\def\c{\gamma} 
\def\d{\delta}
\def\e{\epsilon}           
\def\f{\phi}               
\def\g{\gamma}

\def\j{\psi}

\def\cq{{\cal Q}}

\def\pa{\partial}                              

\def\>{\rangle} 

\def\<{\langle} 
\def\Dsl{D \hskip-.6em \raise1pt\hbox{$ / $ } }

\def\PRD{Phys. Rev. D}

\def\PRL#1#2#3{{\it Phys. Rev. Lett.} {\bf#1} (#2) #3}

\def\NPB#1#2#3{{\it Nucl. Phys.} {\bf B#1} (#2) #3}

\def\PRD#1#2#3{{\it Phys. Rev.} {\bf D#1} (#2) #3}

\def\PLB#1#2#3{{\it Phys. Lett.} {\bf #1B} (#2) #3}

\def\to{\rightarrow}

\def\1ov4{{1\over 4}}

\def\pa{\partial}

\def\xx{\times}

\def\nonu{\nonumber \\{}}

\begin{document}
\begin{flushright}
KUL-TF-98-02 \\
hep-th/9801076
\end{flushright}
\def\titleline{
Branes and anti-de Sitter spacetimes
}
\def\authors{
Harm Jan Boonstra, Bas Peeters,
Kostas Skenderis
}
\def\addresses{
Instituut voor Theoretische Fysica,
KU Leuven, Celestijnenlaan 200D,
3001 Heverlee, Belgium
}
\def\abstracttext{
We consider a series of duality transformations that leads to a
constant shift in the harmonic functions appearing in the description
of a configuration of branes. This way, for several intersections of
branes, we can relate the original brane configuration which is
asymptotically flat to a geometry which is locally isometric to $adS_k
\xx E^l \xx S^m$. These results imply that certain branes are dual to
supersingleton field theories. We also discuss the implications of our
results for supersymmetry enhancement and for 
supergravity theories in diverse dimensions.
}
\makefront

Duality symmetries have played a prominent r\^{o}le during the  
last few years by providing a handle on non-perturbative 
physics. Using a web of duality symmetries
one can connect all known string theories and eleven-dimensional
supergravity, leading to a conjectural master theory, the $M$-theory\cite{wit},
that contains all others as special limits. The $U$-duality group of $M$-theory
compactified on a torus is usually assumed to be the (discretized) version 
of the global symmetry group of the various maximal supergravity theories 
obtained from eleven-dimensional supergravity by toroidal compactification 
in dimensions $d \leq 10$.
There are, however, also duality transformations\cite{hyun, BPS} 
outside this group that change the asymptotic geometry of spacetime 
and connect Poincar\'{e} supergravities and gauged anti-de Sitter 
supergravities. 
These new duality symmetries were instrumental in the recent 
microscopic derivation of the Bekenstein--Hawking entropy 
formula for non-extremal black holes\cite{KSKS}, and 
they will be the topic of this contribution.

We shall show that using these duality transformations
one can map supersymmetric configurations of $M$-branes and $D$-branes 
that are asymptotically flat to configurations that are locally 
isometric to $adS_k \xx E^l \xx S^m$, where $adS_k$ is the $k$-dimensional 
anti-de Sitter space, $E^l$ is the $l$-dimensional Euclidean space and 
$S^m$ is the $m$-dimensional sphere. 
We shall further argue that the worldvolume theories of the membrane ($M2$), 
the fivebrane ($M5$) of $M$-theory, as well as the worldvolume theory 
of the self-dual threebrane ($D3$) of IIB string theory, are dual to 
the supersingleton field theories of $adS_4$, $adS_7$ and  $adS_5$, 
respectively.
 
The basic mechanism that converts an asymptotically flat space
to an anti-de Sitter space is a dualization w.r.t. an isometry that 
is spacelike everywhere, except at spatial infinity, where it becomes 
null (cf. \cite{topch}). This isometry is a combination of
spacelike isometries and 
a timelike isometry. The latter is present in any stationary solution.
Since the isometry involves time its orbits are non-compact. 
We shall therefore consider the following limiting procedure. 
We shall first compactify the time coordinate and then at the end let 
the radius of the time coordinate go to infinity.   
Notice that the time coordinate is naturally compact in the 
anti-de Sitter space that we obtain, after all dualities
have been performed. Thus, taking the infinite
radius limit corresponds to considering the covering space.

The effect of the new duality symmetries is to remove the constant part 
from the harmonic functions appearing in the brane solutions.
We shall first discuss the case of a single extremal brane.
The case of multiple intersections of branes will be dealt with afterwards. 
The results hold also for non-extremal branes\cite{BPS} and non-extremal 
intersections of extremal branes\cite{KSKS}.
We shall consider the case of type II string theory, and we 
shall also lift our results to $M$-theory branes. 
In all cases, the brane under consideration is wrapped on a torus,
i.e. all the worldvolume coordinates are taken to be periodic.
Hence these solutions can be dualized, using an
appropriate chain of (standard) duality transformations, to the plane
wave solution 
\bea \label{wave}
&&ds^2 = - H^{-1} dt^2 + H (dx_1 + H^{-1} dt)^2 + (dx_2^2 + \cdots +
dx_9^2), \nonu
&& B_{01}=0; \ \ e^{-2\f}=1
\eea
where 
\be
H(r)= 1 + \frac{\cq}{r^6}; \ r^2=x^2_2 + \cdots + x^2_9.
\ee
The solution (\ref{wave}) is dual to the fundamental string solution. 
The off-diagonal part in (\ref{wave}) was chosen in such a way that the 
antisymmetric tensor of the dual solution is regular at the 
horizon $r=0$, i.e. the antisymmetric tensor vanishes 
at the horizon\cite{hor1}. 

Once we have reached the plane wave solution we dualize  along the 
isometry generated 
by $\pa /\pa x_1' = \pa/\pa x_1 - (1/2) \pa /\pa t$.\footnote{
More generally, one may dualize w.r.t. 
$\pa /\pa x_1' = (1/a) (\pa/\pa x_1 - (1/2) \pa /\pa t)$, where $a$
is arbitrary. 
This duality yields also (\ref{fund}) but with 
$\cq \to \cq /a^2$, and the radius $R_1$ of the $x_1$ coordinate 
becomes $R_1 a$.}
The norm squared of this vector is $\cq / r^6$. So the isometry is spacelike 
everywhere except at spatial infinity where it becomes null.
The result is a fundamental string solution 
but with zero constant part in the harmonic function, i.e.
\bea
&&ds^2 = \tilde{H}(r)^{-1} (-dt^2 + dx_1^2) + (dx_2^2 + \cdots + dx_9^2), 
\nonumber  \\
&&B_{01} = \tilde{H}(r)^{-1}; \ \ e^{-2\f}=\tilde{H}(r); \ \ 
\tilde{H}(r)= \frac{\cq}{r^6}, \label{fund}
\eea 
where we have dropped the primes. From here one can dualize back to
the original brane configuration. 

It is instructive to decompose the new duality transformation into 
two steps. First, do a coordinate transformation to reach adapted 
coordinates and then perform a standard $T$-duality transformation.
The coordinate transformation, which we shall call the shift transformation,
is given by\footnote{
For the more general dualization mentioned in the previous footnote,
the transformation is $x_1'=ax_1; \  t'= (1/a)(t + (1/2) x_1)$} 
\be \label{shift}
x_1'=x_1; \ \ t'= t + {1 \over 2} x_1
\ee  
This transformation yields (with the primes dropped) 
\be \label{wave1}
ds^2 = - \tilde{H}^{-1} dt^2 + \tilde{H} (dx_1 + \tilde{H}^{-1} dt)^2 + 
(dx_2^2 + \cdots + dx_9^2)
\ee 
i.e. (\ref{wave}) but with the harmonic function $H$
replaced by $\tilde{H}$. Subsequent $T$-duality along $x_1$ yields 
(\ref{fund}). Observe that (\ref{wave}) and (\ref{wave1}) 
are both solutions {\it in the same coordinate system}
(of course, (\ref{wave1}) is a solution in the transformed coordinate 
system since (\ref{wave}) is a solution).

The shift transformation can be applied to more general configurations
of fundamental strings, solitonic fivebranes, $D$-branes, waves and
Kaluza-Klein monopoles. We will make the following assumptions: (i)
the harmonic functions only depend on the overall transverse
coordinates, of which there are at least three, and (ii) the
configurations are built according to the 
intersection rules based on the `no-force' condition (\cite{Tse1} and
references therein).  Together, these requirements imply that there
are at most four independent charges. Moreover, the fraction of
supersymmetry preserved is $1/2^n$ for $n\leq3$ and $1/8$ for $n=4$ if
$n$ is the number of charges.  Then any harmonic function appearing in
the description of such a configuration can be shifted by a
constant. This can be achieved by mapping the `brane' corresponding to
this harmonic function to a wave, using $S$ and $T$-dualities. Then we
can apply the same shift transformation (\ref{shift}) as before to
change the harmonic function. Finally we can map back to the original
configuration using the same chain of duality transformations.

The shift transformations in the harmonic functions are particularly
interesting for the case of brane configurations with some simple
near horizon geometry. Such branes interpolate between this geometry
and Minkowski spacetime at infinity.
In eleven dimensions, the membrane interpolates
between ${\cal M}_{11}$ and $adS_4\times S^7$, whereas the
fivebrane interpolates between ${\cal M}_{11}$ and $adS_7\times
S^4$\cite{GiTo}.
We may dimensionally reduce the $M2$-brane to a fundamental (type IIA)
string,  do the shift procedure as explained before, and lift this
configuration up to eleven dimensions again. This will give a configuration
which is locally isometric to the $adS_4\times S^7$ solution of 
eleven-dimensional supergravity, that is just the $M2$-brane solution 
with the harmonic function not containing the constant term.
Notice that the solution is only locally isometric to the 
$adS_4\times S^7$ solution since the original brane was wrapped 
on $T^2$, so in the final configuration the various coordinates 
carry appropriate global identifications. The compact spacetime 
isometries associated with the torus are needed in order to 
be able to dualize the original $M2$ to a wave. Similar remarks 
apply to all cases that we discuss. The $M5$-brane is
similarly connected to the $adS_7\times S^4$ solution.
In fact, $adS_4\times S^7$ and $adS_7\times S^4$ are known to be
maximally supersymmetric vacua of $d{=}11$ supergravity.
There is one ten-dimensional $p$-brane with similar properties:
the self-dual threebrane. Its near horizon geometry, $adS_5\times S^5$,
is itself a maximally supersymmetric vacuum of type IIB supergravity.

Since after the duality the asymptotic geometry has changed,
the degrees of freedom should organize themselves into 
representations of the appropriate anti-de Sitter group. 
The latter has some representations, the so-called singleton 
representations, that  have no Poincar\'{e} analogue.
They have appeared in studies of spontaneous 
compactifications of eleven-dimensional supergravity 
on spheres. In particular, the fields of the 
supersingleton representation appear as coefficients in the 
harmonic expansion of the eleven-dimensional fields 
on the corresponding sphere. A crucial property is that the singleton
multiplets can be gauged away everywhere, 
except on the boundary of the anti-de Sitter space \cite{gauge}.   
In particular, it has been argued in the past that  
the singleton field theories of $adS_4$, $adS_7$, $adS_5$ and $adS_3$
correspond to membranes \cite{m2}, fivebranes \cite{NST,m5}, self-dual 
threebranes \cite{NST,m5} and strings \cite{f1}, respectively.
It has actually been shown that, in all cases, the world-volume
fields of the corresponding $p$-brane form a supersingleton multiplet.  
Notice also that the anti-de Sitter group $SO(d-1,2)$ coincides 
with the conformal group in one dimension lower. 
Therefore, one ends up with a conformal field theory 
on the boundary. Since we are considering extremal branes,
the theory on the boundary is also supersymmetric. 
It is well-known that the worldvolume theory (in the limit that
gravity decouples) of the membrane 
$M2$, of the fivebrane $M5$, of the self-dual 
threebrane $D3$, as well as of the string,
is a superconformal field theory. We therefore conclude that these branes
are $U$-dual to supersingletons (see also \cite{malda}).

Following these considerations there should exist topological 
field theories in anti-de Sitter spaces that when 
restricted to the boundary yield superconformal field theories.
For the case of $adS_3$, this topological theory 
is simply a Chern-Simons theory. It is indeed well-known that a Chern-Simons 
theory on a manifold with boundary induces a WZW model on this boundary. 
Further evidence for this duality for the case of the $D3$-brane 
was recently given in \cite{FF}. In this work, the boundary of the 
anti-de Sitter space was considered at spatial infinity 
with topology $S^1 \xx S^3$ and only a $U(1)$ multiplet was considered. 
In our case, the boundary should carry the appropriate global identifications,
i.e. it should be $S^1 \xx T^3$ since the original brane was wrapped on 
a torus. It will be interesting to analyze whether the $N{=}4$ 
$SU(k)$ SYM theory on $T^4$ can be extended to a topological theory on 
$adS_5$. Similar remarks apply for the case of $M2$ and $M5$
(for a recent work on the worldvolume theory of $M5$, see \cite{CKP}).

Next let us consider the effect of our duality transformations on 
orthogonal intersections of $M$-branes. The cases of interest are 
the ones with a simple near horizon geometry which is a product containing an
anti-de Sitter spacetime and a sphere. There is only one intersection
of two $M$-branes 
which falls into this class, namely the $M2\perp M5$ solution,
\bea
&&ds^2=H_2^{-\frac{2}{3}}H_5^{-\frac{1}{3}}(-dt^2+dx_1^2)
    +H_2^{-\frac{2}{3}}H_5^{\frac{2}{3}}(dx_2^2)
    +H_2^{\frac{1}{3}}H_5^{-\frac{1}{3}}(dx_3^2+\cdots +dx_6^2)\nonu
    &&\ \ \ \ \ \ 
    +H_2^{\frac{1}{3}}H_5^{\frac{2}{3}}(dx_7^2+\cdots +dx_{10}^2)\,,
\label{25metric}\\
&&F_{r012}= \pm \partial_r H_2^{-1}\ \,,\ \ \ \ \ 
F_{2\a\b\c}=
 \pm \e_{\a\b\c}\partial_r H_5\,,\nonumber
\eea
where the signs differentiate between brane and anti-brane,
and $\e_{\a\b\g}$  is the volume form of the three-sphere surrounding
the intersection.
The harmonic functions are
\be\label{harm}
H_2=1+ {Q_2\over r^2}\ ,\ \ \ \ 
H_5=1+ {Q_5\over r^2}\ ,
\ee
where $r^2=x_7^2+x_8^2+x_9^2+x_{10}^2$. With this choice of
harmonic functions the solution is asymptotically Minkowski.
The near horizon geometry is the geometry for $r\rightarrow 0$. In
this limit the constant parts of the harmonic functions become negligible.
If one takes the harmonic functions (\ref{harm}) without the
constant terms one still has a solution, which
can also be obtained by performing the shift transformation. 
In the latter case, however, we get the solution below plus
a set of global identifications, namely the coordinates $x_i$, 
$i=1, \ldots, 6$ are now all periodic. We get 
\bea
ds^2 &=& Q_2^{-\frac{2}{3}}Q_5^{-\frac{1}{3}}r^2(-dt^2+dx_1^2)
    +Q_2^{-\frac{2}{3}}Q_5^{\frac{2}{3}}(dx_2^2)
    +Q_2^{\frac{1}{3}}Q_5^{-\frac{1}{3}}(dx_3^2+\cdots +dx_6^2)\nonu
&&\ \ \ \ \ \ 
     +Q_2^{\frac{1}{3}}Q_5^{\frac{2}{3}}({1\over r^2}dr^2 +d\Omega_3^2)\,.
\eea
Near the horizon the spacetime factorizes into the product of an
$adS_3$ spacetime (with coordinates $t,x_1,r$), a three-sphere $S^3$
of radius $Q_2^{\frac{1}{6}}Q_5^{\frac{1}{3}}$, and a flat Euclidean
five-dimensional space $E^5$.
Similarly, one finds the other possible orthogonal eleven-dimensional
intersections that give rise to this kind of geometry: for three
charges these are the $M2\perp M2\perp M2$ and
$M5\perp M5\perp M5$ (with three overall transverse directions),
and for four charges the $M2\perp M2\perp M5\perp M5$ intersection.
We tabulate the results of the shift transformation below. The right hand
column denotes the geometry which the dual solution is locally
isometric to. These are also the geometries near the
horizon considered in \cite{KK}. Let us emphasize that the two
geometries coincide
only locally: in order to perform the shift transformation, we need to
periodically identify some of the coordinates.

\begin{center}
\vspace{.2cm}
\begin{tabular}{|c|c|}
\hline
$M2$ & $adS_4\times S^7$\\ \hline
$M5$ & $adS_7\times S^4$\\ \hline
$D3$ & $adS_5\times S^5$\\ \hline
$M2\perp M5$ & $adS_3\times E^5\times S^3$\\ \hline
$M2\perp M2\perp M2$ & $adS_2\times E^6\times S^3$\\ \hline
$M5\perp M5\perp M5$ & $adS_3\times E^6\times S^2$\\ \hline
$M2\perp M2\perp M5\perp M5$ & $adS_2\times E^7 \times S^2$\\
\hline
\end{tabular}

\vspace{.2cm}
{\bf Table 1}
\end{center}
\vspace{.1cm}

The product geometries have the form
$adS_{p+2}\times E^{q}\times S^{9-p-q}$, where $p$ is the spatial
dimension of the intersection, $q$ is the number of relative transverse
coordinates and $9-p-q$ is the number of overall transverse coordinates
minus one.
A wave can be added to the common string of the $M2\perp M5$
and $M5\perp M5\perp M5$ intersections (as well as
to the single $M2$ and $M5$-branes).
It modifies only the $adS$ part of the corresponding geometry.

It is well-known\cite{Gibb} that some solutions exhibit supersymmetry
enhancement at the horizon. For example, the $M2$, $M5$ and $D3$-branes
break one half of supersymmetry, whereas their near horizon geometries
are maximally supersymmetric vacua of $d{=}11$ supergravity.
Some other cases of supersymmetry enhancement for static $p$-brane
solutions in different dimensions are known,
and in all these cases the near horizon geometry contains a factor
$adS_{k}\times S^{m}$.
We find that in fact all solutions of table 1
exhibit supersymmetry enhancement at the horizon.
It turns out \cite{BPS} that the condition for unbroken supersymmetry,
$\d\j_M =0$ where $\j_M$ is the eleven-dimensional gravitino,
in the background of the intersections in table 1, reduces
to the geometric Killing spinor equations on the anti-de Sitter,
Euclidean and spherical factors of the geometry. In the case of the
$M2\perp M5$ intersection there is one additional projection, whereas
for the intersections with three and four charges there are two
projections needed.
As one projection reduces the supersymmetry by a factor one half
and since anti-de Sitter, Euclidean and spherical geometries
all admit the maximal number of Killing spinors, one concludes
that the solutions in the right column of table 1 have double the
amount of supersymmetry as compared to their brane counterparts
in the left column.

Furthermore we observe that, for the configurations in table 1, a
dimensional 
reduction over one or more of the relative transverse directions
will always give rise to lower dimensional solutions which also
exhibit supersymmetry enhancement at the horizon. This is because
reduction over such directions, corresponding to the Euclidean
directions in the shifted solutions in the right column,
cannot interfere with supersymmetry.
Further applications of $T$-duality in the
relative transverse directions or $S$-duality will also preserve
the quality of supersymmetry enhancement.
This way one obtains a large class of solutions exhibiting
supersymmetry enhancement, including all previously known ones,
such as the four and five-dimensional extremal black holes
with nonzero entropy.
There are some other interesting features that all
these solutions have in common. They have regular (i.e. finite)
dilaton at the horizon (or no dilaton in eleven dimensions),
and in the shifted solutions the dilaton is a constant everywhere.
Besides, the antisymmetric field strengths become covariantly
constant in the shifted solutions, as is the case for the
Bertotti-Robinson solution.
Finally, these solutions are all non-singular.

At first sight, it seems that the shift transformation relates two solutions
with different amounts of unbroken supersymmetry.
However, as we mentioned,
the shift transformation involves $T$-duality transformations
along the worldvolume coordinates, which therefore have to be
compact.
Only half of the Killing spinors of anti-de Sitter spacetime
survive compactification of the overall worldvolume directions
as one can immediately verify by looking at the explicit expression
of the Killing spinors\cite{LPT}.
Only locally there is an enhanced supersymmetry.

We conclude by making some comments on the implications of our results
for supergravity theories in various dimensions. Each configuration
in $11$ dimensions with effective geometry of the type $adS_k \xx E^l
\xx S^m$ corresponds to a solution of $11d$ supergravity
with the appropriate amount of supersymmetry. It also follows directly
that, after reduction along $p \leq l$ of the flat directions, the
geometry $adS_k \xx E^{l-p} \xx S^m$ is a solution with the same
amount of supersymmetry (counting the number of spinor components) in
$11-p$ dimensions. In addition we can deduce the existence of solutions
with a certain amount of supersymmetry after spontaneous
compactification on the sphere $S^m$. These compactifications are expected
to give
rise to solutions of gauged supergravities in $11{-}m{-}p$ dimensions
with geometry 
$adS_k \xx E^{l-p}$.
Several of these results are well-known, such as
the spontaneous compactification of $11d$ supergravity on $S^7$,
giving gauged $N{=}8$ supergravity in $d{=}4$, and
the $adS_7\times S^4$ and $adS_5\times S^5$ solutions
of $d{=}11$ supergravity and type IIB supergravity. The anti-de Sitter
parts of these solutions are
maximally supersymmetric vacua of gauged maximal supergravities
in seven and five dimensions\cite{gasu}.

Let us also make some remarks about the $10d$ case. 
The ${\cal M}_7\times S^3$ solution of type I supergravity
corresponding to the fivebrane with shifted harmonic function
must correspond to the $1/2$ supersymmetric ${\cal M}_7$ solution with linear
dilaton of gauged $N{=}2$ $d{=}7$ supergravity
\cite{SalSez}.
Also, the solution of a string in a fivebrane gives rise, after the shift,
to $adS_3\times E^4\times S^3$ geometry, and $adS_3\times
T^4$ is a solution of $N{=}2$ $d{=}7$ gauged supergravity as
well\cite{HKL}.
These observations strongly suggest that $N{=}1$ $d{=}10$ supergravity
compactified on a three-sphere yields 
$N{=}2$ $d{=}7$ gauged supergravity (see also \cite{DTN}).
Similar studies can be made for the other cases.


\end{document}